\documentclass[aps,preprint]{revtex4}
\usepackage{amsfonts}
\usepackage{amsmath}
\usepackage{amssymb}
\usepackage{graphicx}
\usepackage{epstopdf}
\usepackage{epsfig}
\begin{document}
\preprint{ }
\title{Deterministic Dense Coding and Faithful Teleportation with Multipartite Graph States}
\author{Ching-Yu Huang$^{1}$\footnote{896410093@ntnu.edu.tw}, I-Ching Yu$^{1}
$\footnote{896410029@ntnu.edu.tw}, Feng-Li Lin$^{1}$\footnote{linfengli@phy.ntnu.edu.tw}, and Li-Yi Hsu$^{2}$
\footnote{lyhsu@cycu.edu.tw}}
\affiliation{${}^{1}$ Department of Physics, National Taiwan Normal University, Taipei,
116, Taiwan}
\affiliation{${}^{2}$ Department of Physics, Chung Yuan Christian University, Chung-li
32023, Taiwan}

\begin{abstract}
We proposed novel schemes to perform the deterministic dense coding and
faithful teleportation with multipartite graph states. We also find the
sufficient and necessary condition of a viable graph state for the proposed
scheme. That is, for the associated graph, the reduced adjacency matrix of the
Tanner-type subgraph between senders and receivers should be invertible.

\end{abstract}
\maketitle

The discoveries of dense coding and teleportation \cite{te,de}, two impossible
tasks in classical information theory, launched the extensive explorations and
studies on quantum information science. For more than a decade, people have
been searching for the deeper connection between quantum physics and
information science. Wherein, as an intriguing feature of quantum physics,
quantum entanglement has been exploited as the physical resource both in
quantum communication and quantum computation. Different entangled states are
requested for different quantum information processing. For example, to
perform either dense coding or teleportation, the entangled states are always
employed as quantum channel. In particular, the graph states play an eminent
role in many applications of quantum information such as the scalable
measurement-based quantum computation \cite{LC1,g1,g2,g3,g4}, the additive and
non-additive quantum error-correction codes \cite{gott,non}. As for their
physical realization, the graph states can be scalably generated based on the
realistic linear optics \cite{Do}. Recently, the six-photon graph state has
been experimentally demonstrated \cite{six}. Also, it has been reported that
graph states can be also effectively prepared using cavity QED \cite{QED}.

In the original proposals, the deterministic dense coding and faithful
teleportation require two-qubit maximal entangled states, which, in fact, are
equivalent to the simplest connected two-qubit graph states under local
operations. Lee \emph{et. al. }firstly showed the possibility of teleporting
two-qubit state using a four-qubit entangled state \cite{lee}. Rigolin studied
the dense coding and teleportation using multipartite entangled states
\cite{gou1,gou2,gou3}. Recently, Yeo and Chua proposed the schemes of
teleportation and dense coding with a genuine four-qubit entangled state
\cite{YC}, which has been verified equivalent to some specific graph states
under local operations \cite{LE}. In addition, very recently, Wang and Ying
proposed the deterministic distributed dense coding and perfect teleportation
schemes with stabilizer states \cite{WY1,WY2}. Eventually, any stabilizer
state is equivalent to a graph state under local Clifford operations
\cite{lc,lc2,LC2}. Therefore, the graph states are implicitly regarded as the
physical resource for the deterministic dense coding and faithful teleportation.

In this letter, we study the deterministic dense coding and faithful
teleportation with the graph states associated with the simply connected
graphs, which are nontrivial multipartite entangled states. We propose the generalized schemes of the
many-to-one dense coding and the one-to-many teleportation, respectively. In
the proposed scenario, the $2n$-qubit graph state, $\left\vert G\right\rangle
$, which comprises qubits $1$, $2$, $\cdots$, $2n$, is initially prepared. The
associated $2n$-vertex graph is denoted as $\Sigma_{G}=($\emph{V}($\Sigma_{G}%
$), \emph{E}($\Sigma_{G}$)). Therein, the qubit $i$ is associated with the
vertex $i$, and, moreover, each vertex $i$ is associated with a stabilizer
generator, $g_{i}=X_{i}{\displaystyle\prod\limits_{j\in N(i)}}Z_{j}$ (Here we
denote the Pauli matrix $\sigma_{x}$, $\sigma_{y}$, and $\sigma_{z}$ acting on
$i$-th qubit by $X_{i}$, $Y_{i}$ and $Z_{i}$, respectively.). The neighboring
set of vertex $i$ is denoted as $N(i)=$ \{vertex $j|$ ($i$, $j$) $\in$%
\emph{E}($\Sigma_{G}$)\}. The state $\left\vert G\right\rangle $ fulfills the
stabilizer condition $g_{i}\left\vert G\right\rangle =\left\vert
G\right\rangle ,$ $\forall i\in\{1,...,2n\}$. In other words,
\begin{equation}
X_{i}\left\vert G\right\rangle ={\displaystyle\prod\limits_{j\in N(i)}}%
Z_{j}\left\vert G\right\rangle . \label{1}%
\end{equation}
Obviously,
\begin{equation}
\{Z_{i}\text{, }g_{i}\}=0, \label{2}%
\end{equation}
where $\{$ , $\}$ is anti-commutator. In fact, Eqns. (\ref{1}) and (\ref{2})
play a key role throughout.

Before further processing, some notations are introduced here. The vertex set
$V(\Sigma_{G})$ is decomposed as two subsets $V_{S}=\{$vertex $i|1\leq i\leq
n\}$ and $V_{R}=\{$vertex $i|n+1\leq i\leq2n\}$. The edge set \emph{E}%
($\Sigma_{G}$) is decomposed into three subsets \emph{E}$_{SR}=\{(i$, $j)|$for
the edge with$\;$$i\in V_{S}$ and $j\in V_{R}$ $\}$, \emph{E}$_{S}=\{(i$,
$j)|i$, $j\in V_{S}$ $\}$, and \emph{E}$_{R}=\{(i$, $j)|$ $i$, $j\in V_{R}\}$.
The graph is called the Tanner-type graph if \emph{E}$_{S}=\varnothing$ and
\emph{E}$_{R}=\varnothing$. In general, a graph $\Sigma_{G}$ can be decomposed
as $\Sigma_{G}^{T}\oplus\emph{E}_{R}\oplus$\emph{E}$_{S}$, where the
Tanner-type subgraph $\Sigma_{G}^{T}=($\emph{V}($\Sigma_{G}$), \emph{E}%
$_{SR})$.

Here we define the following $n\times n$ sub-ajacency matrices: (i) the
reduced adjacency matrix $\Gamma_{T}$ of $\Sigma_{G}^{T}$, where the entry
$(\Gamma_{T})_{ij}=1$ if $(i,n+j)\in\emph{E}_{SR}$ and $0$ otherwise; (ii) the
adjacency matrix $\Gamma_{S}$, where for $1\leq i$, $j\leq n$ the entry
$(\Gamma_{S})_{ij}=1$ if $(i$, $j)\in$ \emph{E}$_{S}$ and $0$\ otherwise;
(iii) the adjacency matrix $\Gamma_{R}$, where for $n+1\leq i$, $j\leq2n$ the
entry $(\Gamma_{R})_{ij}=1$ if $(i$, $j)\in$ \emph{E}$_{R}$ and $0$%
\ otherwise. Moreover, for later usage it is convenient to define $\Gamma
^{<}_{S,R}$ and $\Gamma^{>}_{S,R}$ which are the lower and upper triangular
part of $\Gamma_{S,R}$, respectively. Note that $(\Gamma^{>}_{S,R})^{T}%
=\Gamma^{<}_{S,R}$ as can be easily seen by definition.

Moreover, given a graph state $|G\rangle$ the set of states
$\{|\overrightarrow{k}\rangle:=|(k_{1},\cdots,k_{2n})\rangle
|{\displaystyle\prod_{i=1}^{2n}}Z_{i}^{k_{i}}|G\rangle\}$ forms an orthogonal
measurement basis so that $g_{i}|\overrightarrow{k}\rangle=(-1)^{k_{i}%
}|\overrightarrow{k}\rangle$. (Hereafter, we will use the short-handed
notation $Z^{\overrightarrow{k}}|G\rangle$ for ${\prod_{i=1}^{2n}}Z_{i}%
^{k_{i}}|G\rangle$)

Now we state the main results of the letter as follows, and then we will
investigate the deterministic dense coding and faithful teleportation in
details, respectively.

\emph{Main results }: For the deterministic dense coding and faithful
teleportation with the graph state $\left\vert G\right\rangle $, the
sufficient and necessary condition is that the reduced adjacency matrix
$\Gamma_{T}$ of the Tanner-type subgraph $\Sigma_{G}^{T}$ must be invertible.

\emph{Deterministic many-to-one dense coding }: In the proposed dense coding
scenario, the qubit $1,\cdots,n$ are distributed among $n$ distant senders,
where the $i$-th sender holds the qubit $i$. The other $n$ qubits
$n+1,\cdots,2n$\emph{ } are held by the receiver, Bob. For clear illustration,
the graph state $\left\vert G^{T}\right\rangle $ with the associated
Tanner-type graph, $\Sigma_{G}^{T}$, is initially prepared. In the encoding
phase, to send Bob two classical bits $a_{i}$ and $b_{i}$, $a_{i}$, $b_{i}$
$\in\{$0, 1$\}$, the $i$-th sender performs the local operation $X_{i}^{a_{i}%
}Z_{i}^{b_{i}}$ on the qubit $i$. Then all qubits at senders' hands are
delivered to Bob. Notably, the encoded state now becomes $\left\vert G_{D}%
^{T}\right\rangle :={\displaystyle\prod_{i=1}^{n}}X_{i}^{a_{i}}Z_{i}^{b_{i}%
}\left\vert G^{T}\right\rangle $. For further procedure, define the $n$-bit
binary message vectors $\overrightarrow{a}$ and $\overrightarrow{b}$ with the
$i$-th components being $a_{i}$ and $b_{i}$, respectively. According to the
(\ref{1}), the encoded graph state can be also written as $\left\vert
G_{D}^{T}\right\rangle ={\displaystyle\prod_{i=1}^{n}}Z_{i}^{b_{i}^{\prime}%
}Z_{n+i}^{a_{n+i}^{\prime}}\left\vert G^{T}\right\rangle $. As a result,
according to the (\ref{2}), $g_{i}\left\vert G_{D}^{T}\right\rangle
=(-1)^{b_{i}^{\prime}}\left\vert G_{D}^{T}\right\rangle $ and $g_{n+i}%
\left\vert G_{D}^{T}\right\rangle =(-1)^{a_{n+i}^{\prime}}\left\vert G_{D}%
^{T}\right\rangle $ $\forall i\in\{1,...,n\}$. Similarly, we define two
$n$-bit binary vectors $\overrightarrow{a^{\prime}}$ and $\overrightarrow
{b^{\prime}}$, where the $i$-th components are $a_{n+i}^{\prime}$ and
$b_{i}^{\prime}$ respectively.

In the decoding phase, Bob is firstly to find all components of
$\overrightarrow{a^{\prime}}$ and $\overrightarrow{b^{\prime}}$. That is, he
is to measure the eigenvalues of all stabilizer generators using quantum
circuits. Such task is analogue to finding the syndromes of the stabilizer
quantum error-correction codes. Or, Bob performs the orthogonal measurement
using the orthogonal basis $\{|\overrightarrow{k}\rangle\}$. In this way,
suppose the post-measurement state is $|\overrightarrow{k}\rangle$, then
$\overrightarrow{k}:=(\overrightarrow{b^{\prime}},$ $\overrightarrow
{a^{\prime}})$.

On the other hand, after some calculation we have
\begin{equation}
\overrightarrow{b}=\overrightarrow{b^{\prime}}, \label{4}%
\end{equation}
and $\overrightarrow{a^{\prime}}=\Gamma_{T}\overrightarrow{a}$, or
equivalently,
\begin{equation}
\overrightarrow{a}=\Gamma_{T}^{-1}\overrightarrow{a^{\prime}}. \label{5}%
\end{equation}
That is, Bob can verify the message vectors $\overrightarrow{b}$ and
$\overrightarrow{a}$ using Eqns. (\ref{4}) and (\ref{5}), respectively.
Notably, to guarantee the deterministic decoding, the map : $\overrightarrow
{a^{\prime}}\rightarrow\overrightarrow{a}$ must be \emph{bijective}
(one-to-one and onto). From (\ref{5}) this requires the reduced adjacency
matrix $\Gamma_{T}$ of $\Sigma_{G}^{T}$ must be invertible.

Now we turn to the deterministic dense coding using the general $2n$-qubit
graph state $\left\vert G\right\rangle $ with non-empty sets $\emph{E}_{R}$
and \emph{E}$_{S}$ of the associated graph $\Sigma_{G}$. Note that, by
definition, the adjacency matrix $\Gamma_{T}$ is irrelevant of $\emph{E}_{R}$
and \emph{E}$_{S}$. That is, the (\ref{5}) is unchanged for any associated
graph $\Sigma_{G}$. With straight calculation, the (\ref{4}) should be revised
as
\begin{equation}
\overrightarrow{b}=\overrightarrow{b^{\prime}}+\Gamma_{S}\overrightarrow{a}.
\label{6}%
\end{equation}
Lastly, the nonempty $E_{R}$ will only complicate the quantum circuits to
extract the eigenvalues $\overrightarrow{a^{\prime}}$ and $\overrightarrow
{b^{\prime}}$, and can be taken care appropriately in designing the decoding
circuits. To sum up, in decoding phase Bob firstly measures the eigenvalues of
stabilizer generators to derive $\overrightarrow{a^{\prime}}$ and
$\overrightarrow{b^{\prime}}$, respectively. Then he decodes the message
vector $\overrightarrow{a}$ using (\ref{5}). Once\ $\overrightarrow{a}$\ is
identified, the map : $\overrightarrow{b^{\prime}}\rightarrow\overrightarrow
{b}$ is also bijective. Finally he can determine the message vector
$\overrightarrow{b}$ using (\ref{6}).

Notably, two graph states $\left\vert G\right\rangle $ and $|{\tilde{G}%
}\rangle$ are locally equivalent if there is a local unitary $U\in\lbrack
U(2)]^{2n},$ such that $|{\tilde{G}}\rangle=U\left\vert G\right\rangle $. The
induced transformation between the corresponding graphs $\Sigma_{G}$ and
$\Sigma_{\tilde{G}}$ is called local complementation \cite{LC1,LC2}. On the
other hand, the capacity of dense coding is invariant under local operation.
Therefore, if $\left\vert G\right\rangle $ can be exploited for the
deterministic dense coding, $|{\tilde{G}}\rangle$ will also do for the same
task. In other words, the invertibility of the reduced adjacency matrix
$\Gamma_{T}$ of the Tanner-type subgraph $\Sigma_{G}^{T}$ is preserved under
local complementation.

\emph{Faithful one-to-many teleportation }: In the proposed teleportation
scenario, the sender, Alice, is to teleport the unknown qubits $i^{\prime}$ to
the $i$-th distant receiver $\forall i\in\{1$, $2$, $\cdots$, $n\}$.
Similarly, the qubit $i^{\prime}$ is associated with the vertex $i^{\prime}$,
and we denote the vertex set as $V_{S^{\prime}}=\{$vertex $i^{\prime}|1\leq
i\leq n\}$. Without loss of generality, the density matrix of the $n$-qubit
unknown state is denoted by $\rho_{u}$ and
\begin{equation}
\rho_{u}=%
%TCIMACRO{\dsum \limits_{z_{1},\cdots z_{n},x_{1},\cdots x_{n},=0}^{1}}%
%BeginExpansion
{\displaystyle\sum\limits_{z_{1},\cdots z_{n},x_{1},\cdots x_{n},=0}^{1}}
%EndExpansion
\lambda_{\overrightarrow{z},\overrightarrow{x}}%
%TCIMACRO{\dprod \limits_{i=1}^{n}}%
%BeginExpansion
{\displaystyle\prod\limits_{i=1}^{n}}
%EndExpansion
Z_{i^{\prime}}^{z_{i}}X_{i^{\prime}}^{x_{i}}, \label{7}%
\end{equation}
where the information about the unknown state is encoded in $\lambda
_{\overrightarrow{z},\overrightarrow{x}}$, and $\overrightarrow{z}%
=(z_{1},\cdots,z_{n})$ and $\overrightarrow{x}$ $=(x_{1},\cdots,x_{n})$.
Hereafter, we will use the short-handed notation $\sum_{\overrightarrow{x}}$
for $\sum_{x_{1},x_{2},\cdots=0}^{1}$.

To achieve the teleportation task, the $2n$-qubit graph state, $\left\vert
G\right\rangle $ is initially prepared as addressed before. Wherein, $n$
qubits $\{n+1,\cdots,2n\}$ are at Alice's hand, and the qubit $\emph{i}\in
V_{S}$, $1\leq i\leq n,$ is held by the $i$-th receiver. Then Alice performs
the $2n$-qubit joint measurement with the orthogonal measurement basis
\footnote{Our measurement base here is more general than the one considered in
\cite{YC} where the measurements on vertices in $V_{R}$ are trivial so that
there it requires more specific 4-qubit entangled state for the faithful
teleportation.}%
\begin{equation}
\{|\overrightarrow{k}\rangle:=|(k_{1},\cdots,k_{2n})\rangle|{\displaystyle}%
%TCIMACRO{\dprod _{i=1}^{n}}%
%BeginExpansion
{\displaystyle\prod_{i=1}^{n}}
%EndExpansion
Z_{i^{\prime}}^{k_{i}}Z_{n+i}^{k_{n+i}}\left\vert G^{\prime}\right\rangle \},
\end{equation}
where the graph state $\left\vert G^{\prime}\right\rangle $ is the $2n$-qubit
state which comprise qubits $\{1^{\prime},\cdots,n^{\prime},n+1,\cdots,2n\}$
and is identical to $\left\vert G\right\rangle $ if\ the qubit $i^{\prime}$ is
replaced by the qubit $i$ $\forall i\in\{1,...,n\}$. Similarly, the
corresponding graph is denoted by $\Sigma_{G^{\prime}}$ and its vertex set is
decomposed as two subsets $V_{S^{\prime}}$ and $V_{R}$. The density matrices
of $\left\vert G\right\rangle $ and $\left\vert G^{\prime}\right\rangle $\ are
denoted as follows by $\rho_{G}$ and $\rho_{G^{\prime}}$ respectively,
\begin{equation}
\rho_{G}:=|G\rangle\langle G|=\frac{1}{2^{2n}}\sum_{\overrightarrow{j}%
}{\displaystyle\prod\limits_{i=1}^{2n}}g_{i}^{j_{i}}\qquad\text{ and }%
\qquad\rho_{G^{\prime}}:=|G^{\prime}\rangle\langle G^{\prime}|=\frac{1}%
{2^{2n}}\sum_{\overrightarrow{j^{\prime}}}{\displaystyle\prod\limits_{i=1}%
^{2n}}(g_{i}^{\prime})^{j_{i}^{\prime}}. \label{8}%
\end{equation}
Wherein, $g_{i}^{\prime}$ is the stabilizer generator of $\left\vert
G^{\prime}\right\rangle $ and can be derived via $g_{i}$ with the local
operator on the qubit $i^{\prime}$ instead of the qubit $i$, where $1\leq
i\leq n$. Since these two corresponding graphs, $\Sigma_{G}$ and
$\Sigma_{G^{\prime}}$ of the states $\left\vert G\right\rangle $ and
$\left\vert G^{\prime}\right\rangle $, are identical except with two different
$n$-vertex label sets ($1,\cdots,n$) and ($1^{\prime},\cdots,n^{\prime}$).
Therefore, these two graphs form a mirror pair with respect to the vertex set
$V_{R}$ as illustrated in Fig. 1, and their associated adjacency matrices such
as $\Gamma_{T}$ and $\Gamma_{T}^{\prime}$ ($\Gamma_{S}$ and $\Gamma
_{S}^{\prime}$) as previously mentioned, are exactly equal.

\begin{figure}[ht]
\center{\epsfig{figure=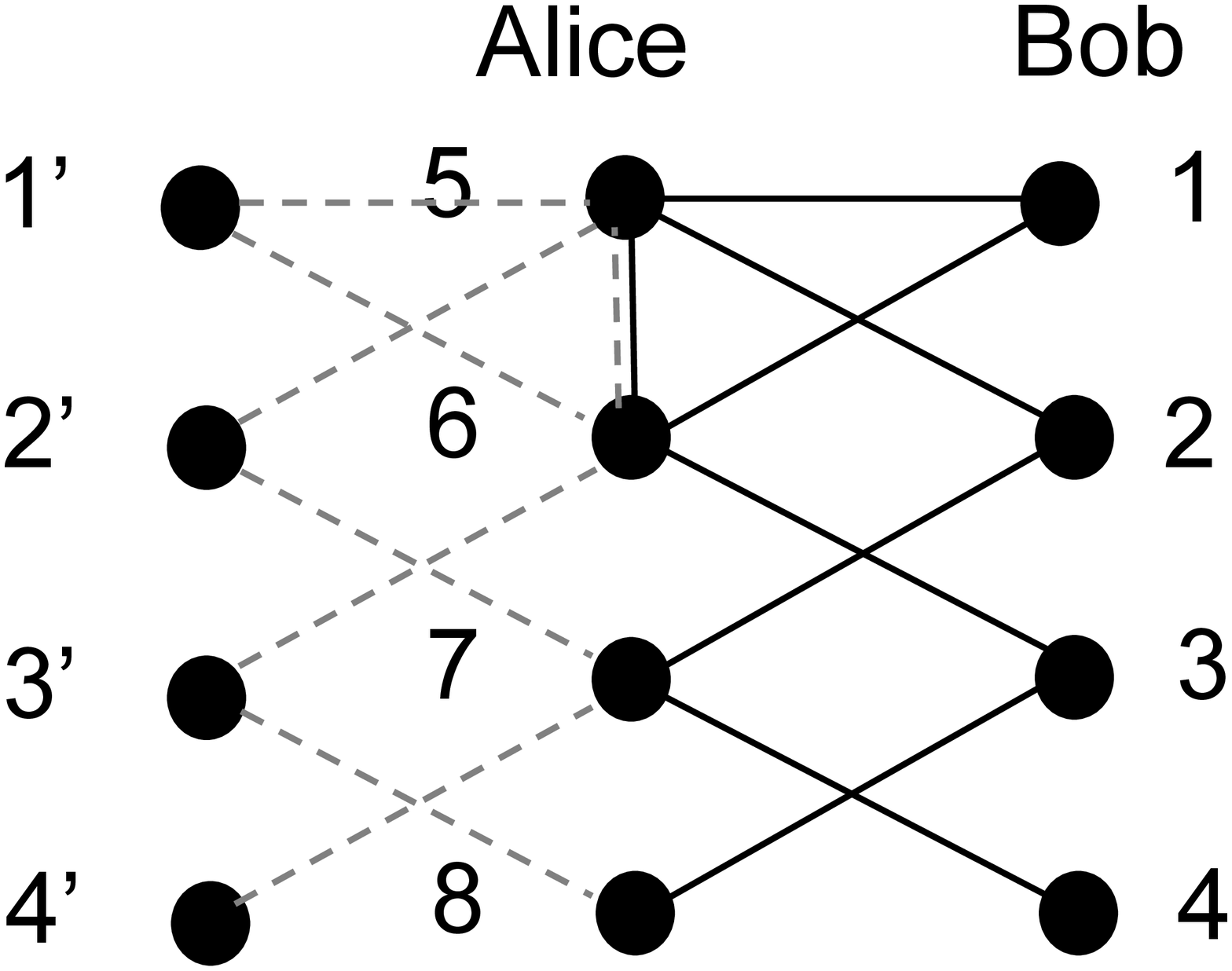,angle=0,width=8cm}}
   \caption{Sketch of a mirror pair of graphs}
   \label{fig1}
\end{figure}  

Let the post-measurement state be $|\overrightarrow{k}\rangle$ after Alice's
measurement, then she announces the $2n$-bit binary vector $\overrightarrow
{k}=(k_{1},k_{2},$ $\cdots,k_{2n})$ of the orthogonal measurement, which will
be exploited to reconstruct the unknown states. Moreover, the corresponding
density matrix of the post-measurement state with qubits $\{1,\cdots,n\}$ held
by $n$ distant receivers is denoted by $\rho_{n}:=tr_{2N}(\rho_{G}\otimes
\rho_{u}\otimes Z^{\overrightarrow{k}}\rho_{G^{\prime}}Z^{\overrightarrow{k}
})$. Here $tr_{2N}$ denotes the tracing-out of qubits $\{1^{\prime}
,\cdots,n^{\prime},$ $n+1,\cdots,2n\}$. After collecting the corresponding
factors we arrive
\begin{align}
\rho_{n}  &  =\sum_{\overrightarrow{x},\overrightarrow{z},\overrightarrow
{j},\overrightarrow{j}^{\prime}}(-1)^{\overrightarrow{x}\; \wedge\;
\overrightarrow{k}_{<}\;+\;\overrightarrow{j}_{>} \;\wedge\;\left[
\overrightarrow{k}_{>} \;+\;\Gamma^{>}_{R}\; (\overrightarrow{j}_{>}
+\overrightarrow{j}^{\prime}_{>})\right]  +\overrightarrow{x}\;\wedge
\;\Gamma^{<}_{S} \;\overrightarrow{j}^{\prime}_{<}\;+\; \overrightarrow{j}%
_{<}\; \wedge\; \Gamma^{<}_{S}\; \overrightarrow{j}_{<}}\;\lambda
_{\overrightarrow{z},\overrightarrow{x}}\nonumber\label{rhon1}\\
&  \times{\displaystyle\prod_{i=1}^{n}}Z_{i}^{\sum_{p\in N(i)}j_{p}}%
X_{i}^{j_{i}} \quad tr_{2N}\left[  Z_{i^{\prime}}^{z_{i}+\sum_{p\in
N(i^{\prime})} j_{p}^{\prime}}X_{i^{\prime}}^{x_{i}+j_{i}^{\prime}}%
Z_{n+i}^{\sum_{p\in N(n+i)}(j_{p}+j_{p}^{\prime})}X_{n+i}^{j_{n+i}%
+j_{n+i}^{\prime}}\right]  .
\end{align}
For conciseness, in the above we have defined $\overrightarrow{j}_{<}%
:=(j_{1},j_{2},\cdots,j_{n})$ and $\overrightarrow{j}_{>}:=(j_{n+1}%
,j_{n+2},\cdots,j_{2n})$, similarly for $k_{<}:=(k_{1},k_{2},\cdots,k_{n})$
and $k_{>}:=(k_{n+1},k_{n+2},\cdots,k_{2n})$. In fact, the expression
(\ref{rhon1}) can be read directly from the graph $\Sigma_{G}$ and
$\Sigma_{G^{\prime}}$. Recall the stabilizer $g_{i}:=X_{i}{\displaystyle\prod
\limits_{j\in N(i)}}Z_{j}$ so that the exponent of $X_{i}(X_{i^{\prime}})$ in
(\ref{rhon1}) is just the sum of $j_{i}(j_{i}^{\prime})$ and $x_{i}$, and
similarly the exponent of $Z_{i}(Z_{i^{\prime}})$ is the sum of $z_{i}$ and
$j_{i}(j_{i}^{\prime})$ of its associated neighboring vertices. In collecting
the above exponents one should interchange $X$ and $Z$ at the same vertex, and
using the fact $Z^{b}X^{a}=(-1)^{a\wedge b}X^{a}Z^{b}$ it results in the phase
factor in Eq (\ref{rhon1}). More specifically, (a) collecting the exponents
associated with vertices in $V_{S^{\prime}}$ yields the phase factor
$(-1)^{\sum_{i=1}^{n} x_{i} \wedge(k_{i} + \sum_{p \in N(i),\; p<i} j^{\prime
}_{p}) }$; (b) collecting the exponents associated with vertices in $V_{R}$
yields the phase factor $(-1)^{\sum_{i=1}^{n} j_{n+i} \wedge(k_{n+i} + \sum_{p
\in N(n+i),\; p>n+i} j^{\prime}_{p} +\sum_{p \in N(n+i),\; p<n+i} j_{p}) }$;
and (c) collecting the exponents associated with vertices in $V_{S}$ yields
the phase factor $(-1)^{\sum_{i=1}^{n} j_{i} \wedge\sum_{p \in N(i),\; p>i}
j_{p} }$. Finally, to arrive the compact form of the phase factor in
(\ref{rhon1}) we have used the following identities%

\begin{equation}
\sum_{p\in N(i),\; p<i}j^{\prime}_{p}=\sum_{q=1}^{n}(\Gamma^{<}_{S})_{i,q}
\;j^{\prime}_{q},\quad\sum_{p\in N(i),\; p>i} j_{p}=\sum_{q=1}^{n} \left[
(\Gamma_{T})_{i,q}\; j_{n+q} + (\Gamma_{S}^{<})_{i,q}\; j_{q}\right]
\label{neigh1}%
\end{equation}
and
\begin{equation}
\sum_{p\in N(n+i),\; p>n+i}j^{\prime}_{p}=\sum_{q=1}^{n}(\Gamma^{>}_{R}%
)_{i,q}\;j^{\prime}_{n+q}, \quad\sum_{p\in N(n+i),\; p<n+i}j_{p}=\sum
_{q=1}^{n}[(\Gamma_{T})_{q ,i}\; j_{q}+(\Gamma^{>}_{R})_{i,q}\; j_{n+q}].  
\end{equation}

Taking the trace in (\ref{rhon1}) results in the following $4n$ Kronecker
deltas
\begin{equation}
\delta(j_{n+i}+j_{n+i}^{\prime}) \label{kk1}%
\end{equation}%
\begin{equation}
\delta(\sum_{p\in N(n+i)}(j_{p}+j_{p}^{\prime})) \label{kk2}%
\end{equation}%
\begin{equation}
\delta(x_{i}+j_{i}^{\prime}) \label{kk3}%
\end{equation}
and
\begin{equation}
\delta(z_{i}+\sum_{p\in N(i^{\prime})}j_{p}^{\prime}) \label{kk4}%
\end{equation}
for $i=1,\cdots,n$. The symmetric form between $j_{p}$ and $j_{p}^{\prime}$ in
(\ref{kk1})-(\ref{kk2}) is due to the fact that graphs $\Sigma_{G}$ and
$\Sigma_{G^{\prime}}$ form a mirror pair. These $4n$ Kronecker deltas will be
used to \emph{completely} eliminate the sum over the dummy vectors
$\overrightarrow{j},\overrightarrow{j^{\prime}}$ in (\ref{rhon1}) if they are
all linearly independent, which is also the condition to guarantee the
faithful teleportation. The sufficient and necessary condition of the $4n$
linearly independent Kronecker deltas turns out to be the same as the one for
the deterministic dense coding, namely, the reduced adjacency matrix
$\Gamma_{T}$($\Gamma_{T}^{\prime}$) of the Tanner-type subgraph $\Sigma
_{G}^{T}$($\Sigma_{G^{\prime}}^{T}$) is invertible. This can be seen as follows.

After imposing the conditions (\ref{kk1}) in the arguments of Eq. (\ref{kk2}),
the $n$ Kronecker deltas (\ref{kk2}) can be reduced to
\begin{equation}
\delta(\sum_{\ell=1}^{n}(\Gamma_{T})_{\ell,i}(j_{\ell}+j_{\ell}^{\prime
}))\label{kk5}%
\end{equation}
which is associated with each vertex $n+i\in V_{R}$. Therefore, the linear
independence of the $n$ Kronecker deltas (\ref{kk2}) is the same as the one
for (\ref{kk5}), which is equivalent to the invertibility of the reduced
adjacency matrix $\Gamma_{T}$. Once the linear independence is guaranteed, it
is easy to see that (\ref{kk1}) and (\ref{kk2}) are reduced to $\delta
^{(2n)}(\overrightarrow{j}+\overrightarrow{j^{\prime}})$, which also yields
the other $2n$ linearly independent Kronecker deltas $\delta(x_{i}+j_{i})$ and
$\delta(z_{i}+\sum_{p\in N(i)}j_{p})$ obtained from (\ref{kk3}) and
(\ref{kk4}). Using the Kronecker deltas $\delta^{(2n)}(\overrightarrow
{j}+\overrightarrow{j^{\prime}})$ and $\delta^{(n)}(\overrightarrow
{x}+\overrightarrow{j})$ one can reduce the phase factor in (\ref{rhon1}) into
$(-1)^{\overrightarrow{x}\;\wedge\;\overrightarrow{k}_{<}\;+\;\overrightarrow
{j}_{>}\;\wedge\;\overrightarrow{k}_{>}}$. Besides, these Kronecker deltas
also help to turn the factor $Z_{i}^{\sum_{p\in N(i)}j_{p}}X_{i}^{j_{i}}$ in
(\ref{rhon1}) into $Z_{i}^{z_{i}}X_{i}^{x_{i}}$. Moreover, using (\ref{neigh1}) and the above
relations the $n$ Kronecker deltas $\delta(z_{i}+\sum_{p\in N(i)}j_{p})$ are
reduced to $\delta^{(n)}(\overrightarrow{j}_{>}+\Gamma_{T}^{-1}%
(\overrightarrow{z}+\Gamma_{S}\overrightarrow{x}))$ which are linearly
independent if $\Gamma_{T}$ is invertible. One can then use these Kronecker
deltas to solve the dummy vector $\overrightarrow{j}_{>}$ in terms of
$\overrightarrow{x}$ and $\overrightarrow{z}$ to reduce the phase factor in
(\ref{rhon1}) further. The phase factor now becomes 
\begin{equation}
(-1)^{\overrightarrow
{x}\wedge\overrightarrow{k}_{<}\;+\;\Gamma_{T}^{-1}(\overrightarrow{z}
+\Gamma_{S}\overrightarrow{x})\wedge\overrightarrow{k}_{>}}=
(-1)^{\overrightarrow{x}\wedge [I+(\Gamma_T^{-1}\Gamma_S)^T]\overrightarrow{k}+
\overrightarrow{z}\wedge (\Gamma_T^{-1})^T\overrightarrow{k}}:=(-1)^{\overrightarrow{x}\wedge\overrightarrow{c_{x}}
\;+\;\overrightarrow{z}\wedge\overrightarrow{c_{z}}}
\end{equation}
Therein, the $i$-th
components of $\overrightarrow{c_{x}}$ and $\overrightarrow{c_{z}}$,
$c_{x,i\text{ }}$and $c_{z,i\text{ }}$are functions of $\overrightarrow{k}$.
As a result, the $i$-th receiver can derive the values of  $c_{x,i\text{ }}%
$and $c_{z,i\text{ }}$since $\overrightarrow{k}$ has been publicly announced by Alice. 

Consequently, as long as $\Gamma_{T}$ is invertible we can arrive
\begin{equation}
\rho_{n}=\sum_{\overrightarrow{x},\overrightarrow{z}}(-1)^{\overrightarrow
{x}\wedge\overrightarrow{c_{x}}\;+\;\overrightarrow{z}\wedge
\overrightarrow{c_{z}}}\;\lambda_{\overrightarrow{z},\overrightarrow{x}%
}{\displaystyle\prod\limits_{i=1}^{n}}Z_{i}^{z_{i}}X_{i}^{x_{i}}.\label{13.5}%
\end{equation}
Finally, in the correction phase, the $i$-th receiver performs $Z_{i}%
^{c_{x,i}}X_{i}^{c_{z,i}}$ on the qubit $i$ to recover $\rho_{u}$ faithfully.

\emph{Discussion}: It should be emphasized that the viable graph for our proposed
scheme should have full rank reduced adjacency matrix $\Gamma_{T}$ of the Tanner sub-graph. 
As an example, in the four-qubit case,
there are two inequivalent graph states. One is the four-qubit
Greenberger-Horne-Zeilinger\ (GHZ) state associated with the star graph, 
the other is the cluster state associated with the linear cluster graph. 
Consequently, the cluster state rather than GHZ state can be
exploited for deterministic dense coding and faithful teleportation.
So is the $2n$-qubit GHZ states which is known to be associated with the star
graph and cannot be exploited in our proposed schemes.  

Moreover, the rank of $\Gamma_T$ is related to the Schmidt measure of the graph state with respect to the 
bi-partition into $V_S$ and $V_R$ \cite{g4}, which is kind of the channel capacity for the quantum communication
between Alice and Bob. This somehow explains why the GHZ state is not viable here since its Schmidt measure is not maximal though
the state itself is maximally entangled.  It is hoped that the above Schmidt-type measure is related to the recently proposed
negative quantum conditional entropy, which indicates the potential to \textquotedblleft receive future quantum
information for free\textquotedblright\ via teleportation and some other ways \cite{na}.

 It is known that the resultant graph under the local complementation (LC) corresponds to the local unitary operation acting
on the original graph state \cite{LC1}, therefore, it will not affect the dense coding and teleportation as
shown in \cite{LE} for the four-qubit case. This also holds true for more general graphs considered in our case, and
can be understood as follows. By definition, the LC on the vertex $i$ is to complement the edges associated with
vertices in $N(i)$. It is then easy to see that the above LC action corresponds to add the column (or row) vector 
associated with vertex $i$ in $\Gamma_T$ to the other columns (or rows) associated with the vertices in $N(i)$. 
Therefore, LC will not change the rank
of $\Gamma_T$, that is, all the LC-equivalent graphs have the same viability for the deterministic dense coding and faithful teleportation
in our proposed scheme.

 Finally, although so far we focus on the two-level multipartite graph states,
the generalization to the dense coding and teleportation with multi-level
graphs states is just straightforward.

We acknowledge the financial support from Taiwan's National
Science Council under Contract No. NSC.96-2112-M-003-014 and NSC.96-2112-M-033-007-MY3.

\end{document}